# Metallization and APPJ treatment of Bismaleimide

A.S.Bhattacharyya[1,3]*, S. Kumar[3], A.Sharma[3], D.Kumar[3], S. B. Patel[4], D.Paul[2], P.P. Dutta[2], G.Bhattacharjee[2]

[1]International Centre for Nanotechnology and Applied Adhesion, Sikkim Manipal Institute of Technology, Sikkim 737136
[2]Department of Mechanical Engineering, Sikkim Manipal Institute of Technology, Sikkim 737136
[3]Centre for Nanotechnology, Central University of Jharkhand, Ranchi: 835205
[4]Centre for Applied Chemistry, Central University of Jharkhand, Ranchi: 835205

*Corresponding author: arnab.bhattacharya@cuj.ac.in

Tel +91 7870674251 (India)

**Abstract**

Bismaleimide (BMI) resins are a new breed of thermosetting resins used mainly for high temperature applications and have major usage in aerospace. FTIR studies have shown the signatures of imide, CNC stretching, malemide and N-H stretching. These BMI polymer coatings were deposited on aluminum and mild steel substrates by sprinkling powers followed by baking. Thermo gravimetric analysis and Differential scanning calorimetric studies showed the degradation temperature of these polymers around 370$^o$C. Aluminum coatings were deposited on BMI previously deposited on Al and mild steel to make a metal-BMI-metal trilayer. These trilayers can solve the problem charging of the aircraft bodies at high altitudes. Atomic force microscopy was done to determine the morphology of the surface. Roughness and thickness measurements of the BMI coatings were carried out by surface profilometer. Vickers microhardness tests showed an increase in hardness of the metal-BMI-metal trilayer. FTIR spectrum showed signature of imides, CNC stretching, maleimide and N-H stretching in BMI. We observed that peak broadens at which shows the release of the stress during thermal treatment of the coating. The coating is subject to variable APPJ conditions which improve the properties at high temperature.

**Keywords:** Polymers, plasma deposition, infrared spectroscopy, surface properties

# 1. Introduction

Bismaleimide (BMI) resins have a significant application in aerospace. They are mainly a subclass of polyimides having properties of dimensional stability, low shrinkage, chemical resistance, fire resistance, good mechanical properties and high resistance against various solvents, acids, and water [1, 2]. Aluminium sheets on the other hand are used extensively in making the spacecraft and aircraft bodies. Coatings of BMI on aluminum sheet will therefore serve as an efficient metal-polymer system which will make the spacecraft perform better in harsh environmental conditions. BMI polymers are thermosetting polymers.

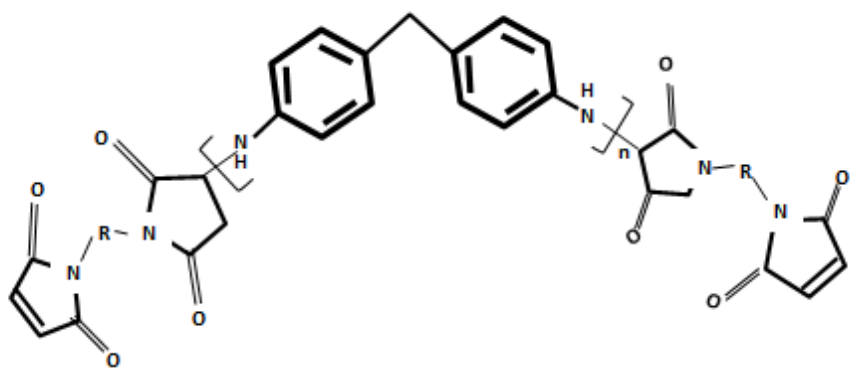

**Fig1**: Chemical structure of BMI [2, 3]

The Bismaleimide (BMI) resins are supplied in powder form and possess a low softening temperature of 90-125°C. These are high temperature BMI polymers have a structure shown in fig 1 [2, 3]. They get cured around 125-150°C but are thermally stable even after 350°C. Aluminum sheets on the other hand have extensive structural advantage as they are light weight yet possess good strength. They are extensively used in the aircraft industry. Aluminium has a melting point of 660°C and thermally stable up to 400°C. Therefore there is a good thermal matching between BMI and aluminium. These resins can be treated thermally to form very good thermoset with excellent thermal, mechanical and chemical properties. They are used for pre-pegs and laminates for the electronic industry, printed circuit boards, diamond wheels and tools, heavy duty insulation and electrical insulation, high temperature applications, glass, aramid and carbon fibre reinforced composite, aerospace and military applications [1, 2, 3]. BMI coating has been used for the corrosion protection of Nd–Fe–B magnets [4]. Thin ceramic coatings Silica coatings has been deposited on bismaleimide (BMI) polymeric substrates as reinforcements or extenders and their Vickers microindentation has been performed which showed cracking[5,6].Siliconized epoxy-phosphorus based bismaleimide coating systems using diglycidylether terminated poly (dimethylsiloxane) (DGTPDMS) and phosphorus-containing bismaleimide (PBMI) as chemical modifiers for epoxy resin has been developed[7].BMI is used to improve the thermomechanical properties of the parent resin[8]. It has been also used for making UV-curable hybrid coatings [9].Studies have also been made on the Environment Effects on Fatigue Life of Carbon/BMI Composite Laminates [10]. Patents have also been filled where BMI coatings have shown potential as a corrosion protection polymer for metals used in vehicles [11]. Studies on electron beam cured BMI coatings have been performed for use as a passivation coating layer for application in microelectronics as a due to their high service temperatures [12]. Addition of BMI also leads to improvement of mechanical properties of epoxy matrices [13].

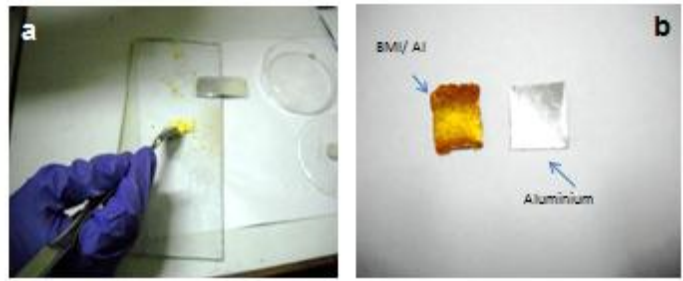

Fig 2: BMI powder (a) sprinkled on aluminum substrates and baked at 100$^o$C to form (b) coatings

## 2. Experiments

The BMI powders were sprinkled on aluminum sheets and put inside oven and thermally treated around 100$^o$C as the softening temperature is between 90-125$^o$C (Fig 2 a). The preparatory steps are shown in fig 3. Atmospheric Pressure plasma jet was applied to the coatings with parameters shown in table 1

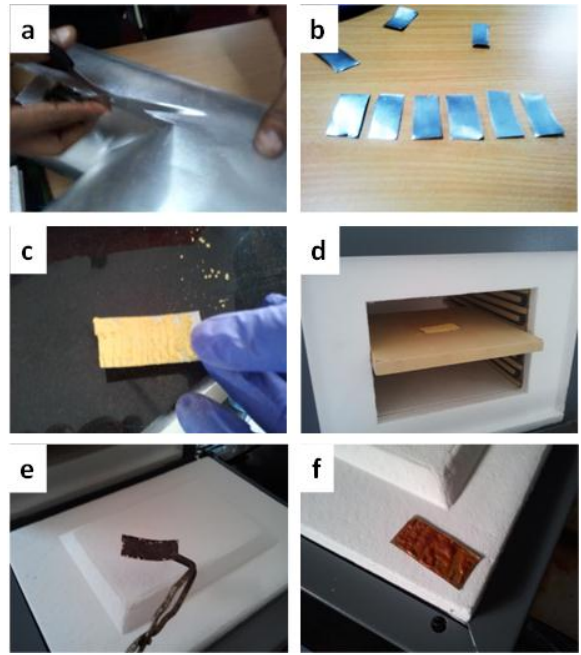

**Fig 3** : Preparation of BMI/Al coatings (a, b) Cutting of Al sheets (c) Sprinkling BMI powders (d) heat treatment in furnace (e, f) formation of coatings

Thermal analysis of the polymer samples were carried out by Thermo Gravimetric Analyser–Differential Scanning Calorimeter supplied by Perkin Elmer, USA. Aluminum coatings were deposited on BMI coated mild steel and aluminum substrates. The coating unit was supplied by Vacuum Equipment Company, India. Talysurf series 2 by Taylor Hobson at National Metallurgical Laboratory (CSIR), Jamshedpur was used for measurement of thickness and roughness of the coating. Vickers microhardness tester supplied by Leica (VMHT Auto), Germany was used to measure the hardness. Surface morphology was studied by Atomic Force Microscopy (AFM) by Nanosurf 2. Atmospheric Pressure Plasma is a low pressure glow discharge. It overcomes the disadvantages of high expense, tedious maintenance and limited sample size associated with normal vacuum plasma unit. The ionized gas generated gets ejected and applied on the sample through a nozzle. The density of plasma in a jet is usually of the order of $10^{11}$-$10^{12}$ free electrons $cm^{-3}$. The APPJ unit established at the Centre for Nanotechnology, CUJ Ranchi used for the surface treatment is shown in Fig 4

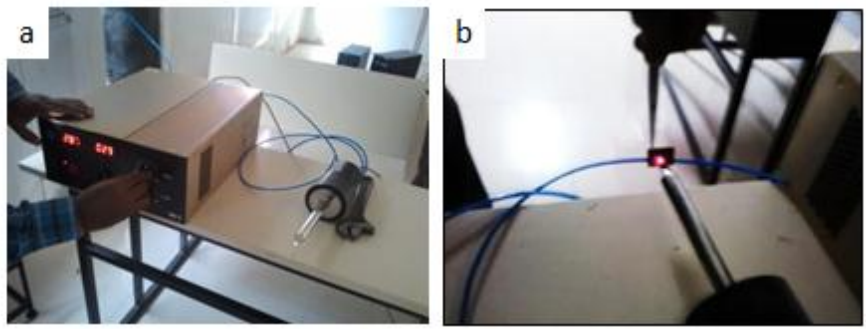

Fig 4: Atmospheric Pressure Plasma Jet Unit

## 3. Results and discussions

FTIR studies (Fig 5 a) of the BMI powder showed signatures of imide with >C=O (1720 cm$^{-1}$) CNC stretching ( 1165 cm-1 and 1390 cm$^{-1}$), malemide and N-H stretching (broad band at 3300 cm$^{-1}$) as per literature [14]. SEM studies of the BMI powders showed particles in the range of 10 to 30 μm (Fig 5 b).

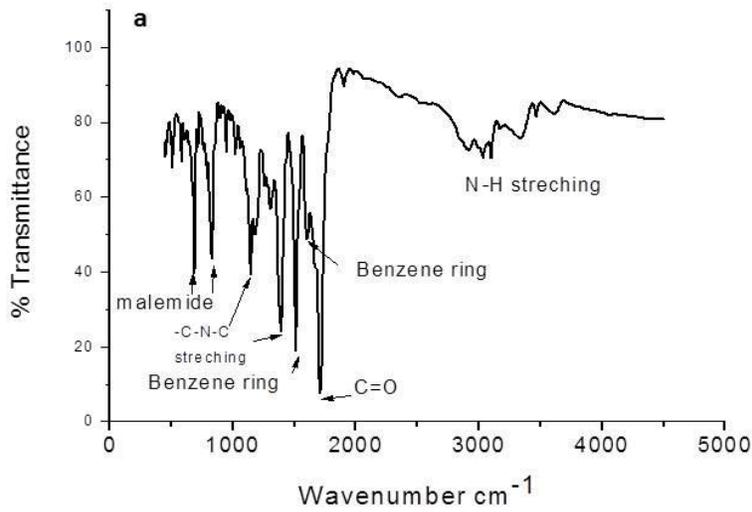

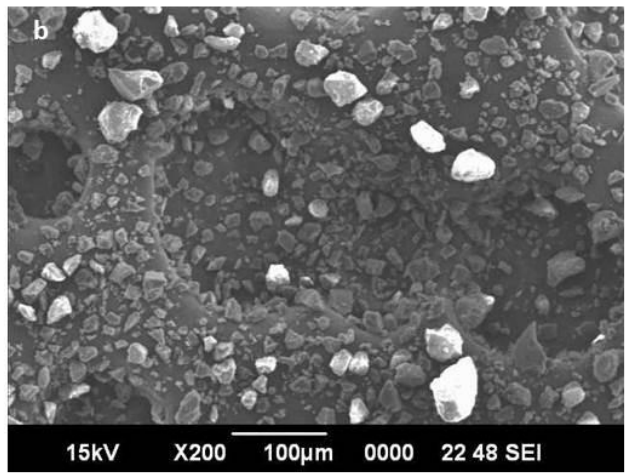

**Fig 5**: (a) FTIR and (b) SEM of BMI powder

The BMI coatings on aluminium were hard and had good adherence to the aluminium substrate as shown in Fig 2(b). The stress induced in the coating was high which caused the coatings to peel of in some cases. The stress arises due to difference in coefficient of thermal expansion between the coating and the substrate. This is a limitation on the part of BMI coatings deposited by baking and needs to be overcome. TGA-DSC studies (Fig 6) showed a thermal decomposition of the BMI at around 370°C which matched with the data available in literature [2].

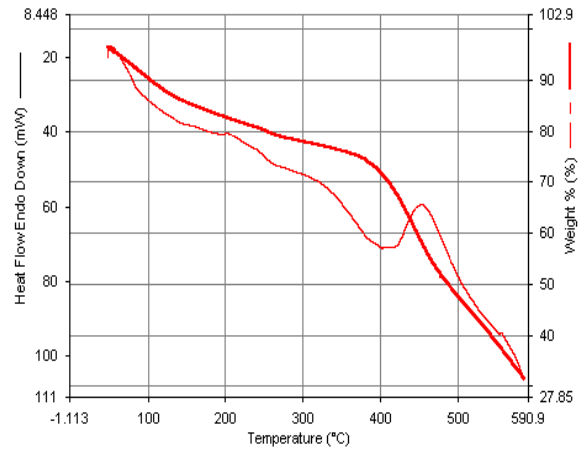

**Fig 6**: TGA-DSC of BMI powders

Bare Al substrate was found to have a roughness of about 100nm. The value increased to 800nm in case of BMI deposited on Al substrates (fig 7). This increase in roughness will result into better adhesive properties of the BMI surface. Aluminum coatings were deposited by the method of resistive heating on the BMI coatings deposited on mild steel and aluminum substrates to form a metal –polymer-metal trilayer as shown in fig 8. Aluminum coatings were also deposited on BMI coated deposited on MS substrates as shown in Fig 9. Microhardness test on aluminum substrate and an Al -BMI-MS trilayer at 25 gf were done. The trilayer also showed a substantial increase in hardness (780 HV). The polymeric materials like BMI which are used in aircraft undergo the problem of charging due to accumulation of secondary electrons. When the electrostatic potential being generated from these accumulated electrons exceeds the dielectric strength of the polymer a breakdown causing damage in the spacecraft can occur. This charging effect can be eliminated if the polymeric surface can be made conducting [15]. Depositing metal like aluminum may be a solution to this problem as aluminum itself has got a large application in aircraft industry. AFM topography of metal /BMI/Mild steel is shown in fig 10. The scanning area was 50 μm × 50 μm. A line profile of the surface is also given below. The surface was found to be smooth with less crests and valleys. The area roughness (Sa) was found to be 52.46 nm$^2$ for a surface area of 2.496 nm$^2$.

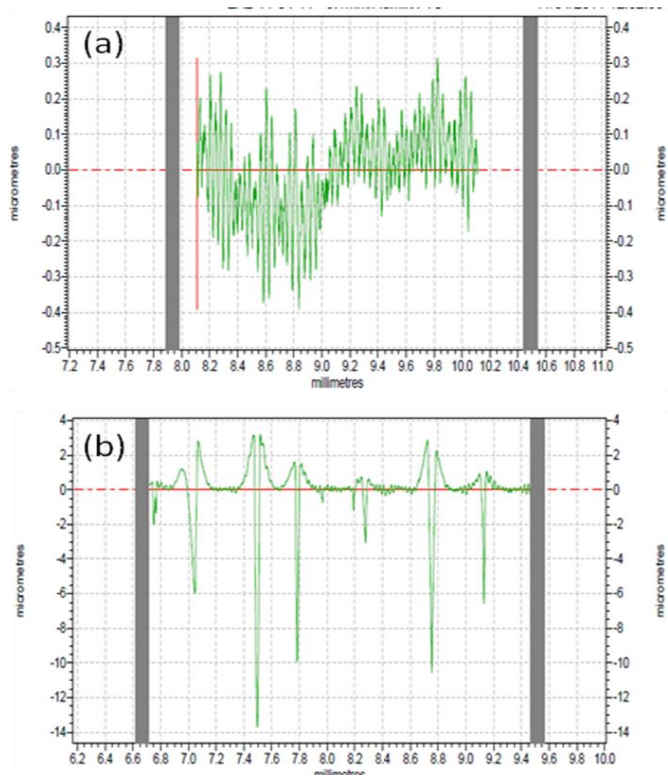

Fig: 7 Roughness profiles of (a) bare al substrate and BMI/Al

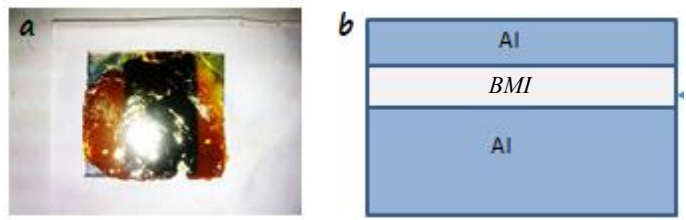

Fig 8: (a) Al-BMI-Al trilayer and (b) its schematic representation

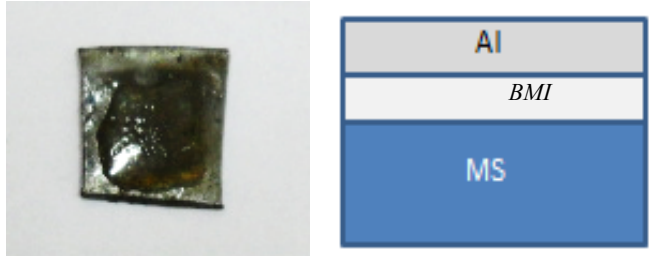

Fig 9: (a) Al-BMI-MS trilayer and its (b) schematic representation

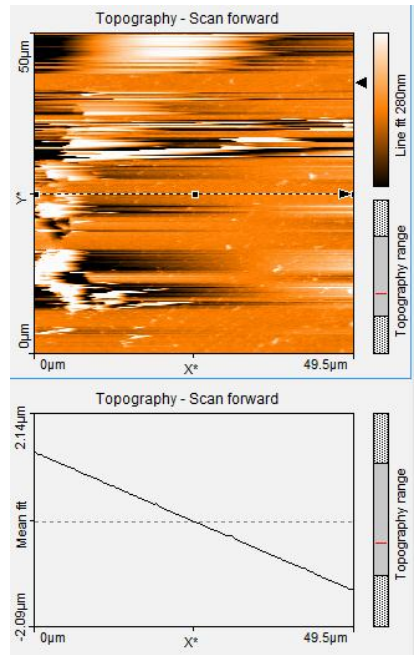

**Fig 10**: Atomic force microscopy of the Al / BMI / MS surface

The FTIR spectra are usually presented as plots of intensity versus wavenumber (in cm$^{-1}$). The intensity can be plotted as the percentage of light transmittance or absorbance at each wavenumber. Quantitative concentration of a compound can be determined from the area under the curve in characteristic regions of the IR spectrum [7]. Concentration calibration is obtained by establishing a standard curve from spectra for known concentrations.

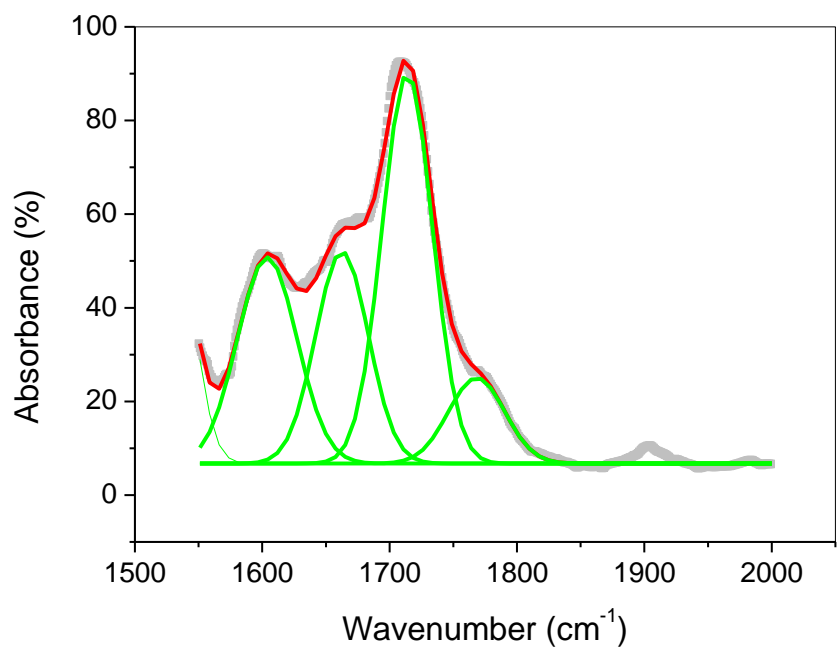

**Fig 11**: Deconvolution of FTIR Spectra of BMI powders from 1550– 2000 cm$^{-1}$

The spectrum was deconvoluted by fitting in Gaussian peaks between 1500–2000 cm$^{-1}$ (Fig 11). Peak at 1768 cm$^{-1}$ obtained was due to C=O whereas at 1713 cm$^{-1}$ the peak corresponded to benzene ring. The peaks corresponding to 1662 cm$^{-1}$ and 1603 cm$^{-1}$ were attributed to N-H and C≡N stretch (Table 1).

Table 1: Parameters obtained after deconvolution of FTIR spectra from 1500 – 2000cm$^{-1}$

| Wavenumber | Area | FWHM | Signature |
|---|---|---|---|
| 1768 | 1014 | 44.2 | C = O |
| 1713 | 4128 | 39.5 | Benzene ring |
| 1662 | 2352 | 41.4 | N-H |
| 1603 | 2582 | 46.8 | C≡N |

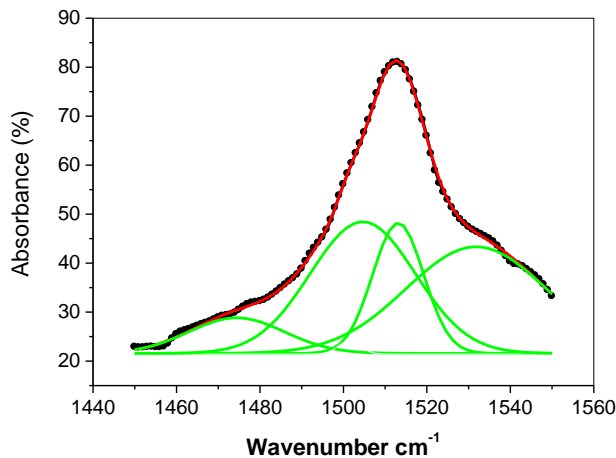

**Fig 12**: Deconvolution of FTIR Spectra of BMI powders from 1450– 1550 cm$^{-1}$

The spectrum was deconvoluted by fitting in Gaussian peaks between 1440 – 1550 cm$^{-1}$ (Fig 12). Peak at 1474 cm$^{-1}$ obtained was due to CH$_2$ bending whereas peak at 1531 cm$^{-1}$ corresponded to benzene ring (Table 2). The N-H bending-stretching and C-N stretching modes are shown schematically below in Fig 13.

Table 3: Parameters obtained after deconvolution of FTIR spectra from 1450 – 1550 cm$^{-1}$

| Wavenumber | Area | FWHM | Signature |
|---|---|---|---|
| 1474 | 209 | 23 | CH$_2$ bending |
| 1504 | 848 | 25 | Non-attributed |
| 1513 | 407 | 12 | Aromatic peak (Benzene ring) |
| 1531 | 917 | 34 | Non-attributed |

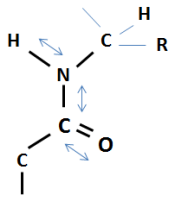

**Fig 13**: The N-H bending-stretching and C-N stretching modes

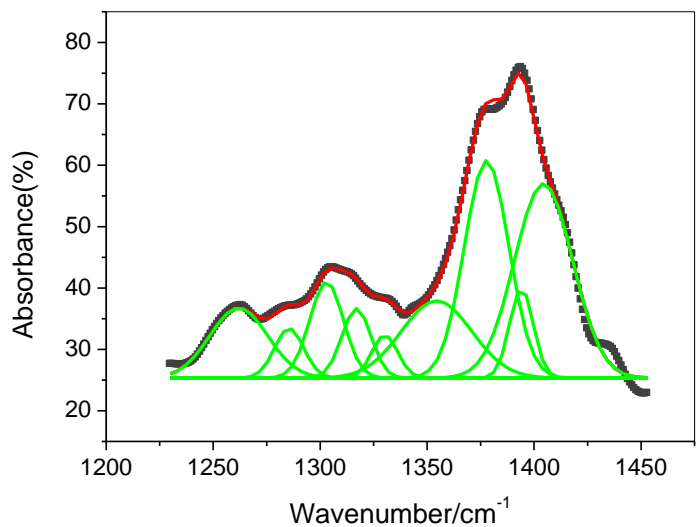

**Fig 14**: Deconvolution of FTIR Spectra of BMI powders from 1225 – 1450 cm$^{-1}$

The spectrum was deconvoluted by fitting in Gaussian peaks between 1225 – 1450 cm$^{-1}$ (Fig 14). Peak at 1404 and 1285 cm$^{-1}$ obtained was due to C-N stretching. The peak occurring at 1378 corresponds to C-CH$_3$ bening whereas at 1262 cm$^{-1}$ the peak corresponded to ether units. (Table 4)

Table 3: Parameters obtained after deconvolution of FTIR spectra from 1225 – 1450 cm$^{-1}$

| Wavenumber | Area | FWHM | Signature |
|---|---|---|---|
| 1404 | 1080 | 27 | C-N stretch |
| 1394 | 203 | 11 | Non-attributed |
| 1378 | 926 | 21 | C-CH$_3$ bending |
| 1354 | 519 | 33 | Non-attributed |
| 1330 | 106 | 12 | Non-attributed |
| 1317 | 193 | 14 | Non-attributed |
| 1303 | 310 | 16 | Non-attributed |
| 1285 | 138 | 14 | C-N stretching |
| 1262 | 386 | 27 | Ether units |

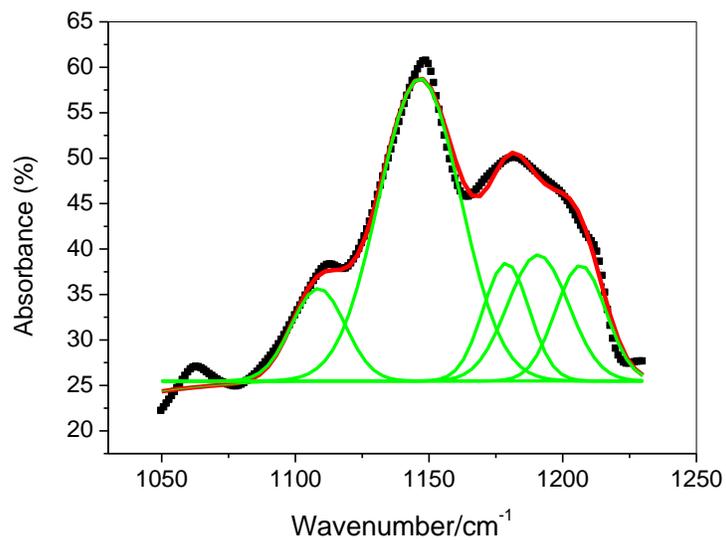

**Fig 15**: Deconvolution of FTIR Spectra of BMI powders from 1050 – 1250 cm$^{-1}$

The spectrum was deconvoluted by fitting in Gaussian peaks between 1050 – 1250 cm$^{-1}$ (Fig 15). Peak at 1147 is attributes to $CH_2$ bending and the peak at 1179 shows C-O-C bending. (Table 4)

Table 4: Parameters obtained after deconvolution of FTIR spectra from 1050 – 1250 cm$^{-1}$

| Wavenumber | Area | FWHM | Signature |
|---|---|---|---|
| 1109 | 256 | 20 | Non-attributed |
| 1147 | 1273 | 31 | $CH_2$ Bending |
| 1179 | 282 | 17 | C-O-C bending |
| 1191 | 397 | 23 | Non-attributed |
| 1207 | 300 | 19 | Non-attributed |

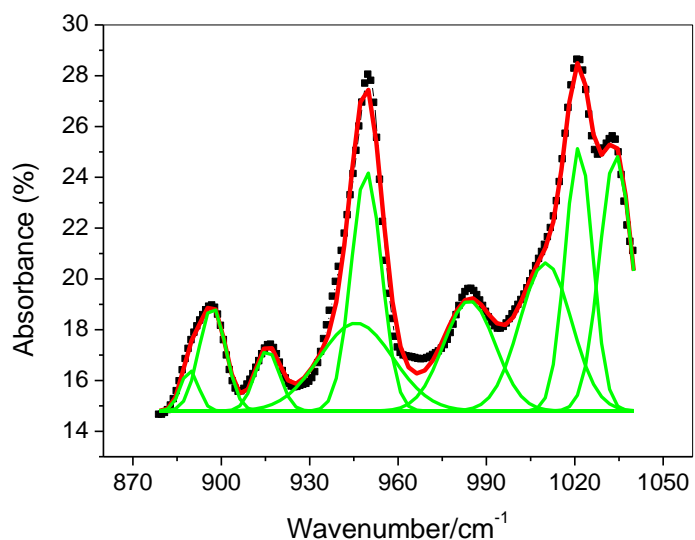

**Fig 16**: Deconvolution of FTIR Spectra of BMI powders from 870 – 1000 cm$^{-1}$

The spectrum was deconvoluted by fitting in Gaussian peaks between 870 – 1000 cm$^{-1}$ (Fig 16). Peak at 950 cm$^{-1}$ obtained was due to $C-CH_3$ bending. (Table 5)

Table 5: Parameters obtained after deconvolution of FTIR spectra from 870 – 1000 cm$^{-1}$

| Wavenumber | Area | FWHM | Signature |
|---|---|---|---|
| 889 | 12 | 6 | Non-attributed |
| 897 | 46 | 9 | Non-attributed |
| 916 | 25 | 8 | Non-attributed |
| 946 | 109 | 25 | Non-attributed |
| 950 | 121 | 10 | C-CH$_3$ bending |
| 984 | 96 | 18 | Non-attributed |
| 1021 | 121 | 9 | Non-attributed |
| 1034 | 139 | 11 | Non-attributed |
| 1010 | 128 | 16 | Non-attributed |

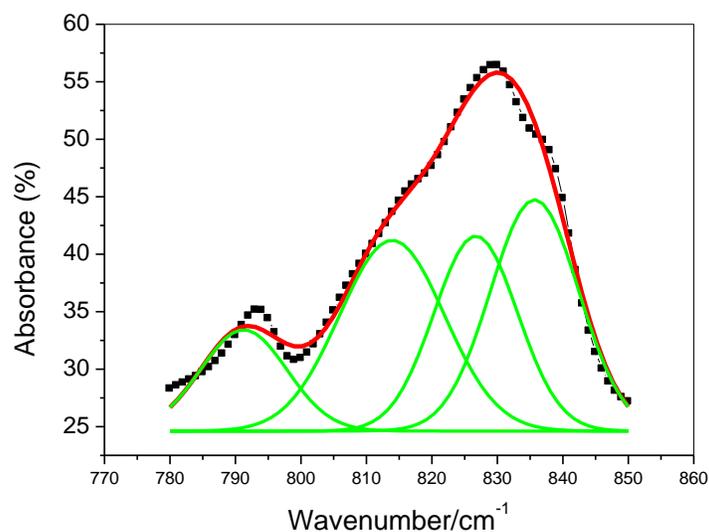

**Fig 17**: Deconvolution of FTIR Spectra of BMI powders from 780 – 850 cm$^{-1}$

The spectrum was deconvoluted by fitting in Gaussian peaks between 780 – 850 cm$^{-1}$ (Fig 17). Peak at 827cm$^{-1}$ occurred due to BMI double bond conversion. NH plane wagging peak occurs at 814 cm$^{-1}$ (Table 6).

Table 7: Parameters obtained after deconvolution of FTIR spectra from 780 – 850 cm$^{-1}$

| Wavenumber | Area | FWHM | Signature |
|---|---|---|---|
| 836 | 340 | 13 | Non-attributed |
| 827 | 272 | 13 | BMI double bond conversion |
| 814 | 334 | 16 | NH Plane wagging |
| 791 | 146 | 13 | Non-attributed |

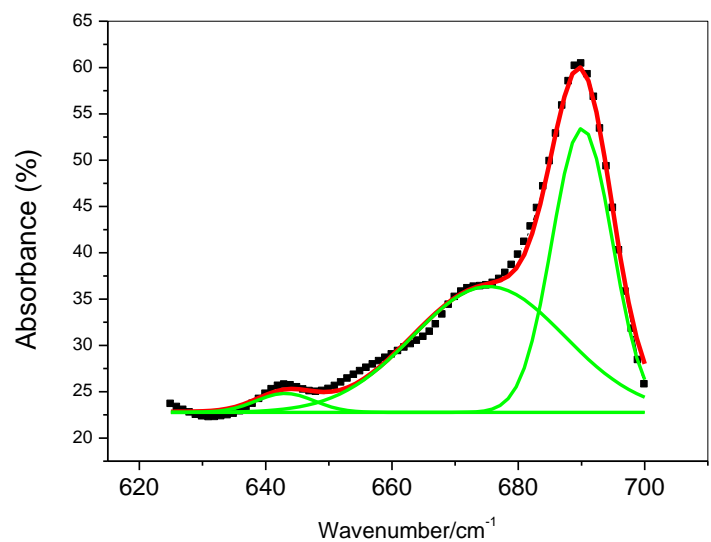

**Fig 18**: Deconvolution of FTIR Spectra of BMI powders from 620 – 700 cm$^{-1}$

Deconvolution of the spectra in the range 620 - 700 cm$^{-1}$ showed peaks at 675 cm$^{-1}$ and 690 cm$^{-1}$ corresponding to OCN bending having a band width of 24 cm$^{-1}$, 10 cm$^{-1}$ and area of 409, 366 a.u (Fig 18). Out of plane N-H bending was evident from peak at 643 cm$^{-1}$ having bandwidth of 9 cm$^{-1}$ and area of 24 a.u (Table 7).

Table 7: Parameters obtained after deconvolution of FTIR spectra from 620 – 700 cm$^{-1}$

| Wavenumber | Area | FWHM | Signature |
|---|---|---|---|
| 690 | 366 | 10 | OCN bending |
| 675 | 409 | 24 | OCN bending |
| 643 | 24 | 9 | NH bending Out of plane |

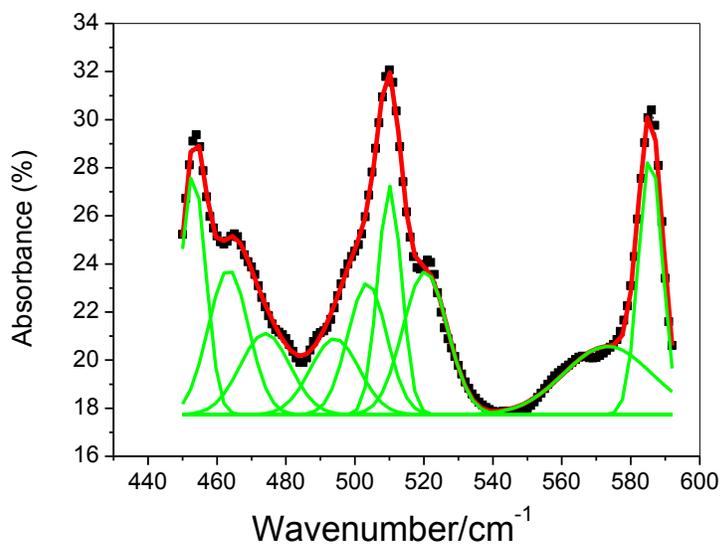

**Fig 19**: Deconvoluted FTIR spectrum between 440 – 600 cm$^{-1}$

Deconvoluting the spectra between 440 – 600 cm$^{-1}$ resulted in numerous peaks (Fig 19) The peaks at 586 cm$^{-1}$ and 573 cm$^{-1}$ were due to out of plane C=O bending. The peaks obtained at lower wavenumbers are attributed to skeletal torsion modes. Due to low energies, these are many populated states. (Table 8)

Table 8: Parameters obtained after deconvolution of FTIR spectra from 440 – 600 cm$^{-1}$

| Wavenumber | Area | FWHM | Signature |
|---|---|---|---|
| 586 | 92 | 7 | Out of plane C=O bending |
| 573 | 90 | 25 | Out of plane C=O bending |
| 520 | 96 | 13 | Out of plane C=O bending |
| 510 | 82 | 7 | Out of plane C=O bending |
| 503 | 77 | 11 | Skeletal distortions |
| 494 | 56 | 14 | Skeletal distortions |
| 473 | 61 | 15 | Skeletal distortions |
| 463 | 88 | 12 | Skeletal distortions |
| 453 | 91 | 7 | Skeletal distortions |

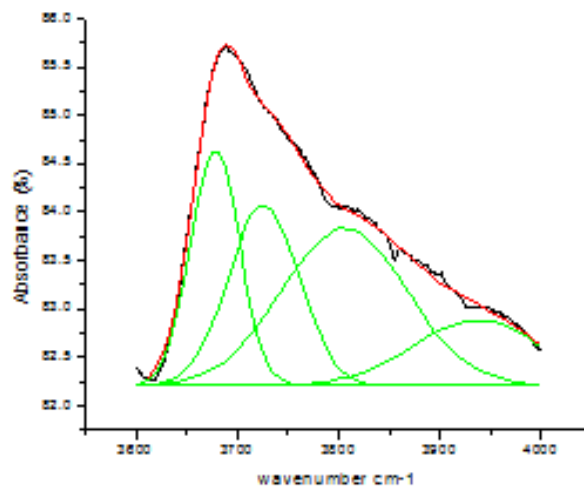

**Fig 20**: Deconvoluted FTIR spectrum between 3600-4000 cm$^{-1}$

Deconvoluting the spectra between 3600 – 4000 cm$^{-1}$ resulted in numerous peaks (Fig 20) This region corresponds to O-H, N-H and C-H stretching. The broad centered at 3726 occurs due to metal OH- stretching vibration [16]. Broadening is due to formation of hydrogen bond (Table 9).

Table 9 : Parameters obtained after deconvolution of FTIR spectra from 3600-4000 cm$^{-1}$

| Wavenumber | Area | FWHM | signature |
|---|---|---|---|
| 3678.1 | 146.94 | 48.281 | Non-attributed |
| 3805.7 | 250.56 | 123.21 | Non-attributed |
| 3726.0 | 162.49 | 69.689 | Metal-OH stretching |
| 3936.5 | 106.93 | 126.45 | Non-attributed |

FTIR spectra of the BMI coatings deposited on Al under different condition are sown in fig 21.

Coatings deposited at higher temperatures showed higher intensity peaks of N-H around 3000 cm$^{-1}$ and others. However the number of prominent peaks reduced and broad peaks were obtained indicating overlapping of vibration frequencies of different bonds at higher temperatures. On applying APPJ the intensity of peaks at higher temperatures got reduced while those at lower temperatures got increased.

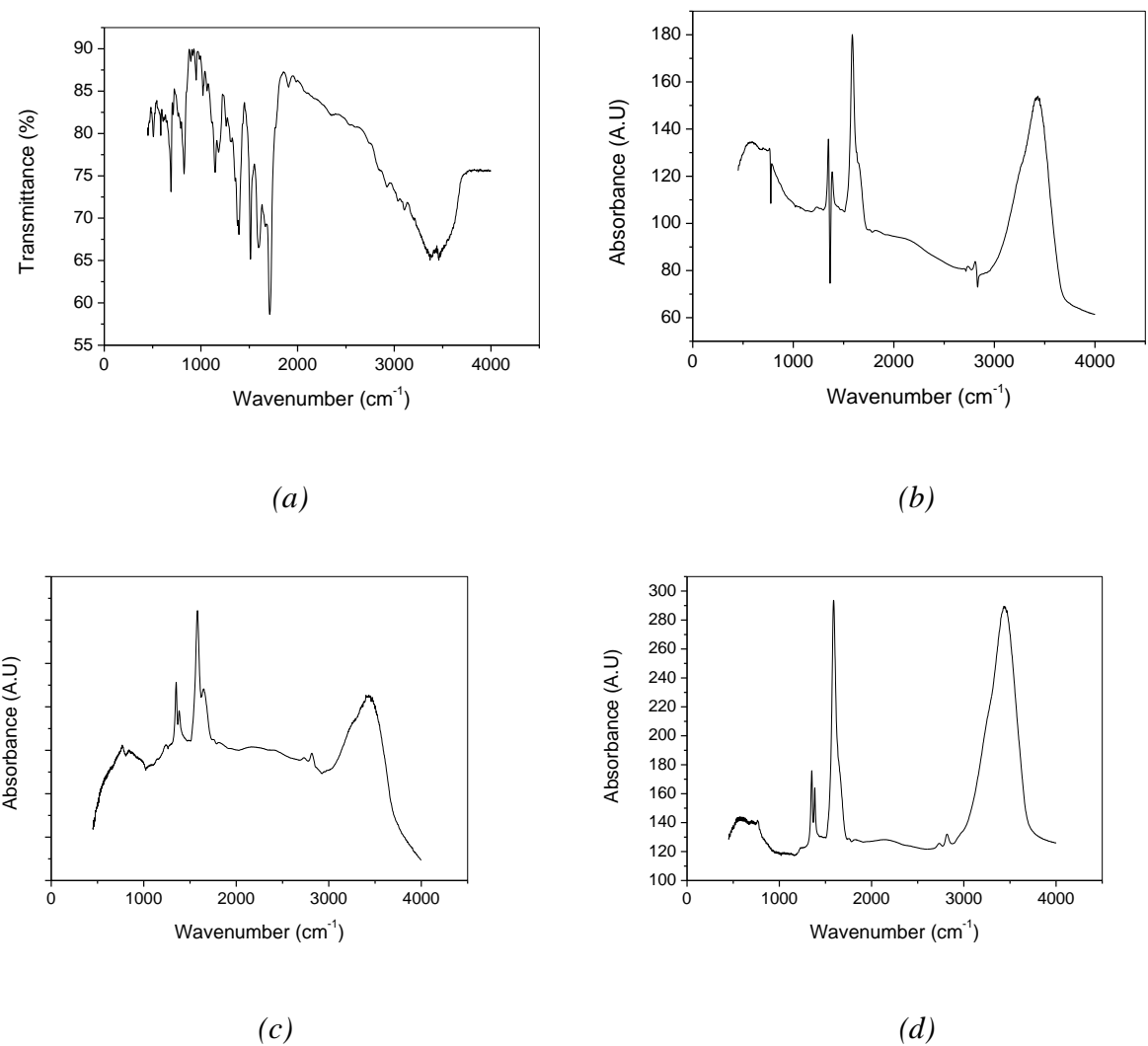

*(a)* *(b)*

*(c)* *(d)*

**Fig 21**: FTIR Spectra of BMI coating cured at (a) 100°C and (b) 200°C (c) 250°C (d) 300°C

The spectra of BMI/Al prepared 300°C (Fig 21 (d)) was deconvoluted into two Gaussian peaks at 3237 cm$^{-1}$ and 3458 cm$^{-1}$ having bandwidth and area of 224 cm$^{-1}$, 15866 and 218 cm$^{-1}$; 41333 for the two peaks respectively(Fig 22).

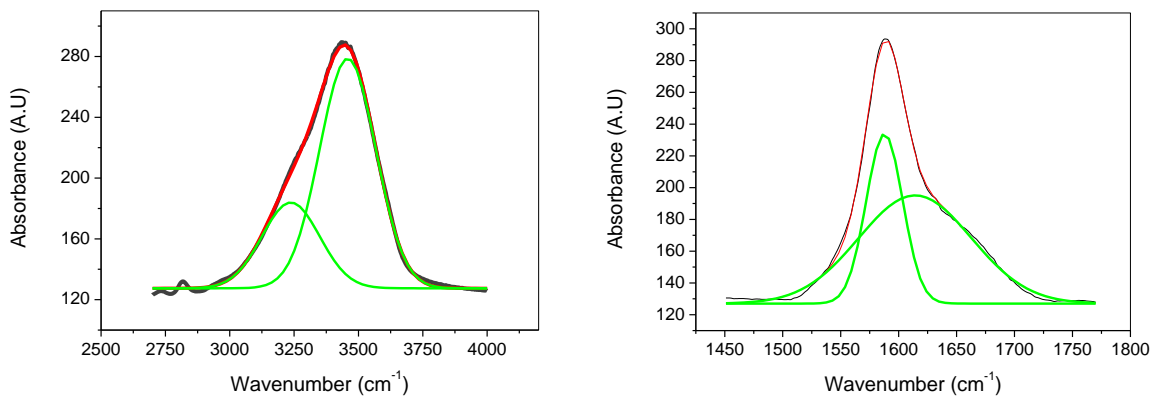

**Fig 21**: Deconvoluted peaks of FTIR spectrum between (a) 2500 – 4000 cm$^{-1}$ and (b) 1450-1750 cm$^{-1}$

Peaks at 1588 cm$^{-1}$ corresponding to N-H bend, 1614 cm$^{-1}$, 3201 cm$^{-1}$ were obtained. The peak obtained at 3435 cm$^{-1}$ was due to amide group. The vibrations occurring in amide are schematically shown in fig 22.

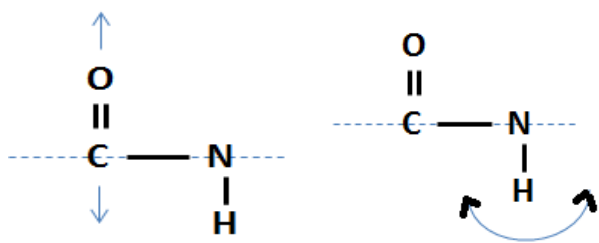

**Fig 22**: Vibrations in Amide

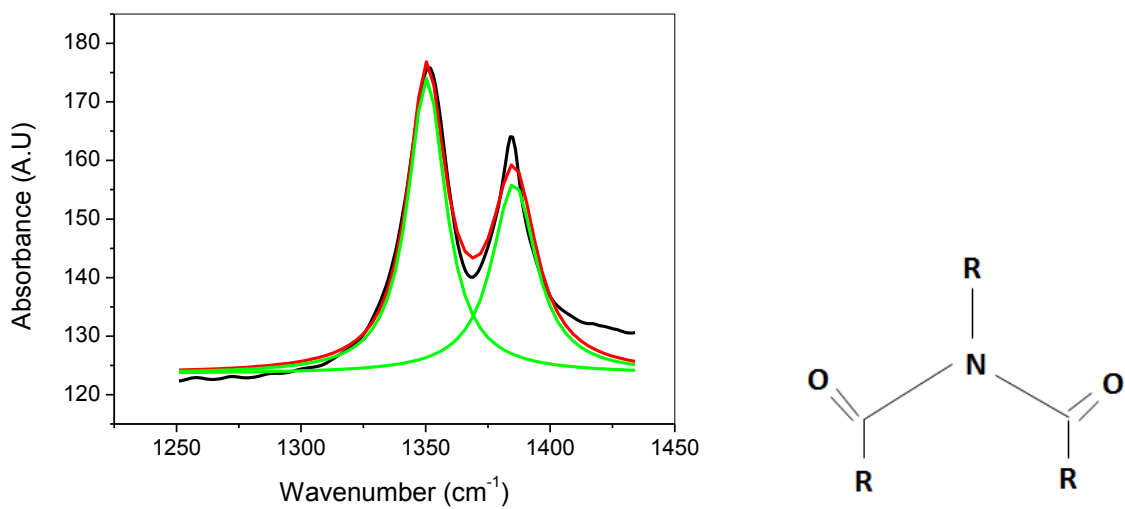

**Fig 23**: (a) FTIR spectrum between 1250 – 1425 cm$^{-1}$ (b) The Imide functional group

For spectrum between 1250 – 1425 cm$^{-1}$, peaks at 1350 cm$^{-1}$ and 1385 cm$^{-1}$ having band width and area 18cm$^{-1}$; 1459 a.u and 22 cm$^{-1}$; 1103 a.u respectively were obtained between 1250 – 1425 cm$^{-1}$ were due to imide characteristic ring vibration shown in Fig 23 (a). The imide functional group consisting of two acyl groups bound to nitrogen is shown schematically in fig 23(b). The SEM image of BMI coatings on Al both pristine and APPJ treated are shown in Fig 24. APPJ is mainly used in industries for activating and cleaning plastic and metal surfaces prior to adhesive bounding and painting process. Application of APPJ in this case has done the surface etching which has caused the pore size to increase and surface looks more rough which will held in better adhesive bonding.

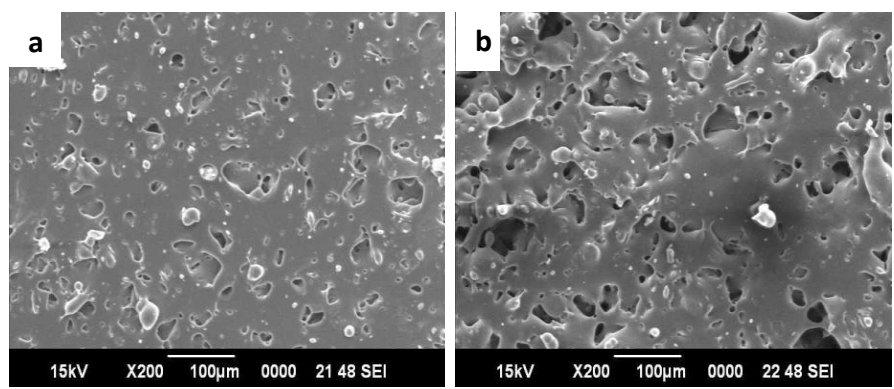

**Fig 24**: SEM of a) BMI/Al and b) APPJ treated BMI/Al coatings

## 4. Conclusions

Bismalemide (BMI) coatings were deposited on aluminium sheets by a simple process of powder sprinkling and baking. The coating were well adhered to the substrate and showed corrosion resistance. TGA-DSC studies showed BMI to be thermally stable upto 370$^o$C. Aluminum was deposited on these BMI coatings by vacuum thermal deposition to make Al/BMI/Al and Al/BMI/MS multilayer which can solve the problem of accumulation of space-charge in aircraft bodies at high altitude. These trilayers showed higher hardness compared to bare Al and MS.AFM gave an insight into the surface morphology of the coatings.

## 4. Acknowledgements

The authors thank Science and Engineering Research Board, India for research grant SERB/F/3482/2012-2013 (Dated 24 September 2012). SEM studies were performed at the Centre for Instrumental facility (CIF), BIT Mesra. The authors also thank Prof. S. Bhowmik for his suggestions and Dr S.K.Mishra, CSIR-NML Jamshedpur profilometry and hardness tests.